\documentclass[aps,prb,unsortedaddress,draft,showpacs,intlimits,amsmath,amssymb,floats,floatfix,twocolumn,10pt]{revtex4}
\usepackage{bm}
\usepackage[final]{graphicx}
\usepackage{ulem}
\usepackage{color}
\usepackage{epsfig}
\usepackage[thinspace]{SIunits}
\usepackage{xspace}
\newcommand{\LSCO}{\mbox{La$_\mathrm{2-x}$Sr$_\mathrm{x}$CuO$_4$}\xspace}
\newcommand{\NCCO}{\mbox{Nd$_\mathrm{2-x}$Ce$_\mathrm{x}$CuO$_4$}\xspace}

\newcommand{\Atwog}{\mbox{$A_{2g}$}\xspace}
\newcommand{\Boneg}{\mbox{$B_{1g}$}\xspace}
\newcommand{\Btwog}{\mbox{$B_{2g}$}\xspace}

\begin{document}
\title{An Investigation of Particle-Hole Asymmetry in the Cuprates via Electronic Raman Scattering}

\date{\today}

\author{B. Moritz$^{1,2}$, S. Johnston$^{3}$, T. P. Devereaux$^{1}$, B. Muschler$^{4}$, W. Prestel$^{4}$, R. Hackl$^{4}$, M. Lambacher$^{4}$, A. Erb$^{4}$, Seiki Komiya$^{5}$, Yoichi Ando$^{6}$}
\address{$^{1}$Stanford Institute for Materials and Energy Science, SLAC National Accelerator Laboratory, 2575 Sand Hill Road, Menlo Park, CA 94025, USA}
\address{$^{2}$Department of Physics and Astrophysics, University of North Dakota, Grand Forks, ND, 58202, USA}
\address{$^{3}$Leibniz-Institute for Solid State and Materials Research Dresden, D-01171 Dresden, Germany}
\address{$^{4}$Walther-Mei\ss{}ner-Institut, Bayerische Akademie der Wissenschaften, 85748 Garching, Germany}
\address{$^{5}$Central Research Institute of the Electric Power Industry, Komae, Tokyo 201-8511, Japan}
\address{$^{6}$Institute of Scientific and Industrial Research, Osaka University, Ibaraki, Osaka 567-0047, Japan}

\begin{abstract}
In this paper we examine the effects of electron-hole asymmetry as a consequence of strong correlations on the electronic Raman scattering in the normal 
state of copper oxide high temperature superconductors. Using determinant quantum Monte Carlo simulations of the single-band Hubbard model, we construct 
the electronic Raman response from single particle Green's functions and explore the differences in the spectra for electron and hole doping away from 
half filling. The theoretical results are compared to new and existing Raman scattering experiments on hole-doped \LSCO and electron-doped \NCCO.  These findings suggest that the Hubbard model with fixed interaction strength qualitatively captures the doping and temperature dependence of the Raman spectra for both electron and hole doped systems, indicating that the Hubbard parameter $U$ does not need to be doping dependent to capture the essence of this asymmetry.
\end{abstract}
\maketitle

\section{Introduction}
The parent compounds of cuprate high temperature superconductors are antiferromagnetic Mott insulators at half filling.~\cite{Lee_Nagaosa_Wen}  By removing electrons from or adding electrons to the CuO$_2$ planes by chemical substitution the antiferromagnetism is suppressed and superconductivity appears over a limited doping range.~\cite{Tallon} At first glance one would expect that the doping leads to effects more or less symmetric around half filling similar to the recently discovered FeAs superconductors.~\cite{DCJohnston} However, in the cuprates the differences originating from either electron or hole doping can be quite significant:~\cite{Hanke_EPJst,Anderson_condmat,Anderson_Ong,Ando_PRL_2004,RMP,Dagan,NPArmitage}  the maximal superconducting transition temperature $T_c$ hardly exceeds 30\,K for electron-doped cuprates while reaching 150\,K for hole-doped materials; in the normal state, while the approximately linear variation with $T$ of the resistivity over wide temperature ranges is a hallmark of the hole-doped systems there is much more doping dependence on the electron-doped side where the resistivity crosses over to a nearly $T^2$ behavior already slightly above optimal doping close to $x=0.15$; in the Raman spectra of hole-doped systems, the \Btwog response is essentially universal over the entire doping range of the superconducting dome whereas, concomitant with the resistivity, the Raman spectra of electron-doped systems changes rapidly and, at low temperatures, exhibits Fermi liquid-like shapes for $x\ge0.16$.

Nevertheless, this asymmetry is not entirely unexpected for strongly correlated copper oxides.~\cite{Emery,Joe,Lee_Nagaosa_Wen} In the insulator at half filling, the wavefunction is composed largely of a superposition of a Cu $d^9$ hole on each 3$d_{x^2-y^2}$ orbital, with a minority of $d^{10}L$ character, wherein the Cu orbitals are filled and a hole occupies the oxygen $2p$ ligand $L$. When hole-doped away from half filling, a $d^9L$ state forms wherein the additional hole gains delocalization energy as well as magnetic exchange energy by occupying the oxygen ligand to form a so-called Zhang-Rice 
singlet.~\cite{Zhang_Rice,Sawatzky}  In contrast doped electrons tend to reside solely on Cu $d^{10}$ sites. This asymmetry can be revealed by comparing 
photoemission with inverse photoemission, or via angle-resolved photoemission spectroscopy (ARPES) in hole- and electron-doped cuprates.~\cite{ARPES_RMP,NPArmitage} More recently the issue of particle-hole asymmetry has been well explored in scanning tunneling microscopy (STM) studies.~\cite{Randeria,OFischer,SCDavis,MGomes} 

In this paper we explore how particle-hole asymmetry can be viewed from Raman scattering measurements. In particular, we are motivated to explore the question 
of whether the low energy particle-hole excitations emerge from doping a Mott insulator while preserving the strength of correlations in the undoped parent in 
the form of Hubbard $U$, or if these excitations are better described in terms of a strongly doping-dependent Hubbard $U$ leading naturally to a collapse of 
the Mott gap not driven by simply adding particles, but by a strong decrease in $U$ with doping.  Recent comparisons of spectral weight transfer observed with 
ARPES, x-ray absorption, and optical spectra have been interpreted in terms of a doping dependent $U$.~\cite{Markiewicz} This differs from the conclusion reached from many other studies of the Hubbard model.~\cite{Millis,Preuss:1995,Eskes_Meinders_Sawatzky,Castellani}

We construct the electronic Raman response using single-particle propagators determined from determinant quantum Monte Carlo simulations of the single-band Hubbard model.  The theoretical results that highlight differences in the spectra for hole and electron doping are compared to results of Raman scattering experiments on \LSCO and \NCCO.  This qualitative comparison suggests that a constant Hubbard interaction $U$ captures the essence of the doping and temperature dependence of the particle-hole asymmetry.

In Section~\ref{sec:theory} we present a brief description of the theoretical calculation and the main results.  Section~\ref{sec:experiments} provides details on the sample preparation, the experimental methods and results for comparison to theoretical calculations.  A discussion of qualitative similarities and differences between experiment and theory appears in Section~\ref{sec:discussion} including comparisons between the extracted scattering rates (Raman resistivity) and the evolution of the Raman spectral weight with doping and temperature.  Finally, we present conclusions in Section~\ref{sec:conclusions}.  


\section{Theory}\label{sec:theory}
\subsection{Model}
The single-band Hubbard Hamiltonian represents an effective low-energy model for the cuprates.~\cite{Zhang_Rice,Anderson}  Its applicability derives from
down-folding models explicitly incorporating planar copper and oxygen degrees of freedom.  Written in a second-quantized real-space representation, the 
Hamiltonian takes the form
\begin{eqnarray}
H&=&-\sum_{ij,\sigma}t_{ij}c^{\dag}_{i,\sigma}c_{j,\sigma}-\mu\sum_{i,\sigma}n_{i,\sigma}\nonumber\\
 &&+\,U\sum_{i}(n_{i,\uparrow}-\frac{1}{2})(n_{i,\downarrow}-\frac{1}{2}).\label{eq:hubbard}
\end{eqnarray}
The operators $c^{\dag}_{i,\sigma}$ and $c_{i,\sigma}$ create or annihilate an electron with spin $\sigma$ at site $i$, respectively, and
$n_{i,\sigma} = c^{\dag}_{i,\sigma}c_{i,\sigma}$ in each spin channel.  The non-zero tight-binding coefficients $\{t_{ij}\}$, restricted to nearest-neighbor
$t$ and next-nearest-neighbor $t'$ hopping, together with the chemical potential $\mu$, that controls the electron filling, define the noninteracting
bandstructure and the Hubbard repulsion $U$ controls the strength of electron-electron correlations.

While this Hamiltonian appears rather simple, it resists an analytical solution in two-dimensions, applicable to the cuprates, and is challenging to solve
numerically, especially for the intermediate range of interaction strengths $U$ believed to be most appropriate to the cuprate problem.  We choose to work with
$U=8t$, equal to the noninteracting bandwidth $W$, that represents a canonical value for the interaction strength in the cuprates related to the
charge-transfer energy between copper and oxygen orbitals in these systems; it also sets the largest energy scale for the problem that can be observed 
directly in the high energy Raman response.  Throughout the theoretical analysis, the nearest-neighbor hopping $t$ serves as the energy unit of the 
problem and we substitute a reasonable estimate for this down-folded hopping integral only for the purpose of comparison to experimental results.

We numerically investigate the Hamiltonian of Eq.~(\ref{eq:hubbard}) using determinant quantum Monte Carlo (DQMC),~\cite{DQMC_1,DQMC_2} an auxiliary-field
technique.  In principle, this method is numerically exact and allows one to determine both single- and multi-particle response functions at finite temperature.
However, computational costs limit investigations to finite-size, small clusters (in either real- or momentum-space) and the fermion sign problem~\cite{Loh_Sign,Troyer_Sign} limits the lowest accessible temperatures where one can still obtain reasonably accurate results.  The finite-size clusters used in this study are larger than those that can be accessed using exact diagonalization and provide a sufficient sampling for reconstructing details of the single-particle self-energy assumed to vary slowly as a function of momentum for the chosen parameters.

The DQMC method supplies the finite temperature, imaginary time propagator $G_{ij}(\tau)$ on a finite-size cluster with periodic boundary conditions. Individual
Markov chains of the Monte Carlo process provide input for the maximum entropy method (MEM)~\cite{Jarrell_Guber_MEM,Alex_MEM} used to Wick rotate the imaginary 
time data to real frequencies using Bayesian inference from separate estimates of the propagator assumed to have a Gaussian statistical distribution over 
different chains.  The data are characterized to ensure that they reasonably satisfy this assumption; and they are preprocessed and/or more data are gathered
to satisfy these conditions.  From the real-space statistical ensemble $\{G_{ij}(\tau)\}$, a discrete Fourier transform yields $\{G_{K}(\tau)\}$ from which
MEM returns the single-particle spectral function $A(\mathbf{K},\omega)$ on the corresponding discrete momentum grid. Once obtained in this fashion, the
single-particle self-energy $\Sigma(\mathbf{K},\omega)$ can be extracted using Dyson's equation and the bare bandstructure corresponding to the tight-binding
model parameters.  Assuming a weak momentum dependence to the self-energy, an interpolation routine provides the value of $\Sigma(\mathbf{k},\omega)$ at an
arbitrary point $\mathbf{k}$ in the Brillouin zone (BZ) and Dyson's equation can be employed to compute $A(\mathbf{k},\omega)$ at that point.

Two-particle response functions such as the charge and spin susceptibility (the dynamic structure factors $S(q,\Omega)$ or optical conductivity $\sigma(\Omega)$) can be evaluated in imaginary time and analytically continued to real frequency using a similar prescription to that followed for the 
single-particle Green's function.~\cite{Jarrell_Guber_MEM}  These quantities satisfy well-defined sum rules that make redefining the spectral functions in 
terms of probability distributions and subsequently normalizing the imaginary time data relatively straightforward.  While in principle bounded, the Raman response does not satisfy any similar sum rule; and considering the significant fermion sign problem that already complicates the analytic continuation by 
adding an additional source of noise and covariance in the data, rather than evaluate the Raman response for different symmetries directly in imaginary time 
(or Matsubara frequency), we evaluate the single-particle Green's function in imaginary time, Wick rotate using MEM, and then estimate the Raman response as~\cite{RMP}
\begin{eqnarray}
\chi^{''}_{\mu}(\Omega) &=& \frac{2}{V\pi}\sum_{k}\gamma^{2}_{\mu}(\mathbf{k})\int^{\infty}_{-\infty}G^{''}(\mathbf{k},\omega)G^{''}(\mathbf{k},\omega+\Omega)\nonumber\\
&&\times\left[f(\omega)-f(\omega+\Omega)\right]d\omega.\label{eq:raman}
\end{eqnarray}
In practice the integral over real frequencies is evaluated numerically using Riemannian integration and the upper and lower limits of the integral are
cut-off at frequencies $\ge 5\,t$ beyond the ``step-edges'' set by the difference in Fermi functions appearing in the integrand, providing sufficient accuracy 
over the studied temperature interval.  The vertices $\gamma_{\mu}(\mathbf{k})$ are chosen to correspond to \Boneg [$\gamma_{\scriptscriptstyle{B_{1g}}}(\mathbf{k})=\frac{1}{2}(\cos(k_{x})-\cos(k_{y}))$] and \Btwog
[$\gamma_{\scriptscriptstyle{B_{2g}}}(\mathbf{k})=\sin(k_{x})\sin(k_{y})$] symmetries that highlight the anti-nodal and nodal portions of the Fermi surface, respectively.  While this method neglects vertex corrections in the Raman response, it captures features that correlate with different intra- and interband charge excitations within the model that qualitatively compare to results from experiments on the cuprates.  The relatively high temperatures, lack of vertex corrections, 
and the simplified Hamiltonian mean that other low energy features that can be seen in the experiment like multi-magnon excitations and phonon degrees of freedom are missing from this analysis.  While the appearance of multi-magnon excitations in the Raman response is usually attributed to the effects of higher-order resonant diagrams off-resonance,~\cite{Loudon-Fleury,Shraiman-Shastry} even two-particle vertex renormalization should contribute to the appearance of these features in the response.~\cite{Chunjing_in_prep}

We use $64$-site square clusters with periodic boundary conditions corresponding to a momentum space grid $\{K\}$ with spacing $\pi/4$ in each direction.
The imaginary time interval has been partitioned into $L=48$ ``slices" of size $\Delta\tau = \beta/L$ running from $0$ to $\beta$.  As noted $t$ serves as
the energy unit of the problem.  For this study, $\beta$ varies between $1/t$ and $3/t$ giving a value of $\Delta\tau$ that varies between $1/48t$ and $1/16t$, controlling the Trotter error while maintaining a reasonable computational time to completion, with the majority of results shown for $\beta=3/t$.


\subsection{Results}

\begin{figure}[t]
\centering
\includegraphics[width=0.45\textwidth]{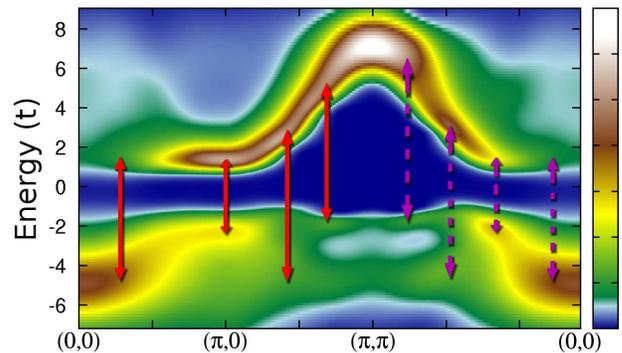}
\caption{(Color online.) Theoretical band dispersion along high symmetry directions for the single-band
Hubbard model at half-filling with parameters $t'=-0.3t$, $\mu=0.0t$, $U=8.0t$, and $\beta=3.0/t$
obtained using determinant quantum Monte Carlo as described in the main text.  The red (solid) arrows
highlight possible $q=(0,0)$ transitions in the \Boneg Raman scattering channel while the purple (dashed) arrows
highlight possible transitions in the \Btwog channel along these high symmetry cuts.  Adapted from Ref.~\onlinecite{Moritz_anomaly_2}.}\label{fig:half-filled_disp}
\end{figure}

Before exploring the Raman response for different regions of parameter space for the Hamiltonian of Eq.~(\ref{eq:hubbard}), let us first look at the
single-particle spectral function for the half-filled model to understand the nature of interband charge excitations that can appear in the response
function.  Figure \ref{fig:half-filled_disp} displays the calculated bandstructure of the half-filled single-band Hubbard model with parameters
$t'=-0.30t$, $\mu=0.00t$, $\beta=3.0/t$ along high symmetry directions in the BZ.  Immediately noticeable are the incoherent lower and upper Hubbard bands
(LHB and UHB) centered near the $\Gamma$-point and $(\pi,\pi)$, respectively.  Above the Fermi level, the UHB has a dispersing branch along the
$(\pi,0)-(\pi,\pi)$ and $(0,0)-(\pi,\pi)$ directions.  Along the $(0,0)-(\pi,0)$ direction, this feature is nearly dispersionless.  In particular, there is
significant spectral weight in the region near $(\pi,0)$ at binding energies near $2\,t$.  Below the Fermi level, while the bulk of the LHB spectral weight is
concentrated near the $\Gamma$-point, there is a dispersing branch along the $(0,0)-(\pi,0)$ and $(0,0)-(\pi,\pi)$ directions that appears to be most
pronounced at binding energies near $-2\,t$.  These features are qualitatively similar to those observed in experiment~\cite{Ronning_anomaly} where a
dispersive feature near $(\pi/2,\pi/2)$ crosses-over to the higher energy valence band, assumed to have significant oxygen character.  
These dispersing features in the LHB and UHB are precursors to a quasiparticle-like band crossing the Fermi level that appears upon either hole or electron doping.

The vertical double-headed arrows (red (solid) in the \Boneg channel and purple (dashed) in the \Btwog channel) that appear in Fig.~\ref{fig:half-filled_disp} mark the energy scale of possible interband charge excitations that can be observed in the various Raman scattering channels.  
Note that the Raman \Boneg and \Btwog vertices highlight the anti-nodal and nodal regions of the BZ, respectively, and, by symmetry, are identically zero along certain high symmetry directions as revealed in the form of each vertex entering Eq.~(\ref{eq:raman}).  
The lowest energy scale for each symmetry is associated with the insulating Mott gap with an onset energy $\sim 2\,t$ and a weak tail at lower energies due to the relatively high temperature.  The main peak associated with the interband transition 
across the Mott gap should occur in both channels at an energy $\sim 4\,t$ with a further prominent transition between the LHB and dispersing tail of the UHB (or precursor to the quasiparticle-like band) at energies between $\sim 6\,t$ and $\sim 10\,t$, although this could be fairly broad.


\begin{figure}[t]
\centering
\includegraphics[width=0.45\textwidth]{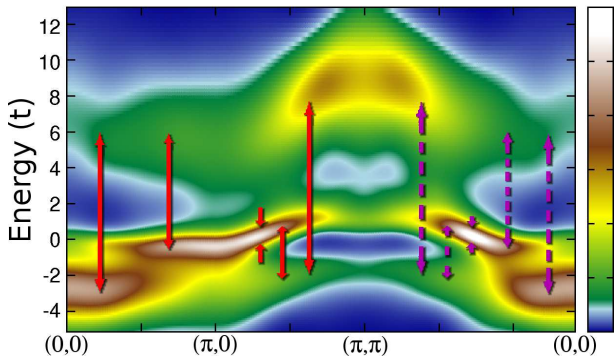}
\caption{(Color online). Theoretical band dispersion along high symmetry directions for the single-band
Hubbard model near optimal ($\sim 15\%$) hole-doping ($t'=-0.3t$, $\mu=-2.5t$, $U=8.0t$, and
$\beta=3.0/t$) with red (solid) and purple (dashed) arrows highlighting the prominent transitions in the 
Raman response in \Boneg and \Btwog channels, respectively.}\label{fig:h-doped_disp}
\end{figure}


\begin{figure}[t]
\centering
\includegraphics[width=0.45\textwidth]{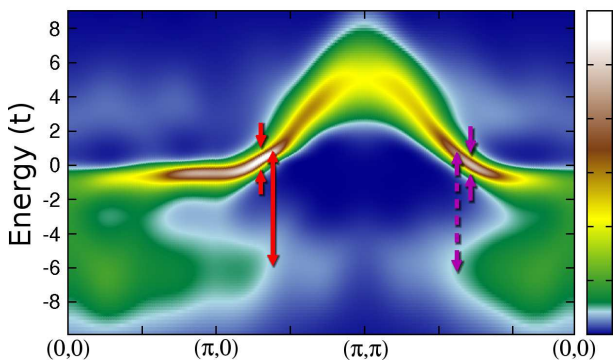}
\caption{(Color online).  Theoretical band dispersion along high symmetry directions for the single-band
Hubbard model near optimal ($\sim 15\%$) electron doping ($t'=-0.3t$, $\mu=2.0t$, $U=8.0t$,
and $\beta=3.0/t$) with red (solid) and purple (dashed) arrows highlighting the prominent 
transitions in the \Boneg and \Btwog Raman response, respectively.}\label{fig:e-doped_disp}
\end{figure}

Upon either hole or electron doping, the gap at the Fermi level closes and the chemical potential
moves into the dispersive portions of either the LHB or UHB forming a
quasiparticle-like band (QPB) at low binding energies near the Fermi level.  This would be reflected in the Raman
response by an onset directly at zero energy and significant low energy (quasiparticle) spectral weight. 

Figure \ref{fig:h-doped_disp} shows the calculated bandstructure
for the hole-doped single-band Hubbard model near optimal doping ($\sim 15\%$) with parameters $t'=-0.3t$, $\mu=-2.5t$, and $\beta=3.0/t$, again along high
symmetry directions in the BZ.  
The QPB here is obviously well separated from the LHB given the significant coexistence of both features along the
$(0,0)-(\pi,\pi)$ and, even more so, the $(0,0)-(\pi,0)$ directions.  The QPB crosses the Fermi level near $(\pi/2,\pi/2)$ and at $\sim (\pi/4,\pi/4)$ the spectral intensity drops, demarcating a cross-over between the QPB and the LHB.  On the whole, the evolution of the QPB qualitatively agrees with the results of ARPES experiments on hole-doped compounds,~\cite{Non_anomaly,Graf_anomaly,Zhou_anomaly,Valla_anomaly,Borisenko_anomaly} including the evolution of spectral intensity and changes in momentum space position and robustness of this ``waterfall"-like appearance as a function of momentum as found in previous work on the single-particle bandstructure.~\cite{Preuss:1995,Hanke_2,DMFT,LDA+DMFT,Alex_Waterfall,Moritz_anomaly,Moritz_anomaly_2}

Figure~\ref{fig:e-doped_disp} shows the spectral function for an electron-doped system with model parameters $t'=-0.3t$, $\mu=2.0t$, and $\beta=3.0/t$ near
optimal electron-doping ($\sim 15\%$).  The LHB, centered at $\sim -6\,t$, has been reduced in intensity from spectral weight transfer into the QPB which
now disperses down across the Fermi level from the precursor in the UHB.  The QPB reaches approximately twice as far below the Fermi level than the QPB under hole-doping.  This dichotomy or asymmetry in the band dispersion between hole and electron-doped systems can be traced back to differences in the shift of the chemical potential either into the LHB or the UHB with doping and the character of the state that then disperses across the Fermi level.~\cite{Moritz_anomaly, Moritz_anomaly_2}

In the hole-doped system the lowest energy scale detected in the Raman response should reflect intraband transitions within the QPB close to the Fermi level at an energy near $1\,t$.  This energy scale also partially reflects the fairly high temperature of this study and would likely decrease with reduction in the model temperature.  An interband transition between the LHB and QPB occurs between $\sim 2\,t$ and $3\,t$ with yet a higher energy transition possible between the LHB and UHB at energies between $\sim 6\,t$ and $9\,t$.  The features in the Raman response should all be fairly broad, not only because of the high simulation temperature, but also because of the intrinsically broad incoherent LHB and UHB that has a diminished spectral weight due to transfers into the QPB.  The transitions that should be prominent in the response are indicated by red (solid) and purple (dashed) arrows in Fig.~\ref{fig:h-doped_disp} for the \Boneg and \Btwog scattering channels, respectively.

In the electron-doped system the general reduction in the LHB spectral weight and the lack of significant dispersive portion of the UHB at higher energies above the Fermi level reduce the number of prominent transitions.
The double-headed arrows in Fig.~\ref{fig:e-doped_disp} indicate two prominent transitions: one from intraband transitions at low energies under $1\,t$ and an additional one from interband transitions between the weak incoherent LHB and dispersive QPB near $7\,t$.  A weak tail exists in the UHB at higher energies; therefore, interband transitions from the QPB or LHB to the UHB also would be weak.


\begin{figure}[t]
\centering
\includegraphics[width=0.45\textwidth]{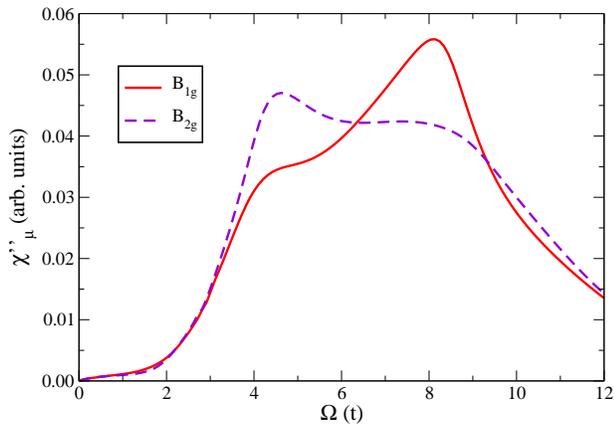}
\caption{(Color online). Theoretical Raman response in \Boneg (red, solid curve) and \Btwog (purple, dashed curve) symmetries for the
half-filled single-band Hubbard model.  (Parameters as in Fig.~\ref{fig:half-filled_disp}).}\label{fig:half-filled_Raman}
\end{figure}

Figure~\ref{fig:half-filled_Raman} shows the response calculated using Eq.~(\ref{eq:raman}) in both the \Boneg and \Btwog scattering channels for the half-filled single-band Hubbard model.  Both channels show an onset at $\sim 2\,t$ corresponding to the Mott gap present in the half-filled model with a weak tail 
at lower energy, due to finite temperature effects, that should shrink at lower temperatures.  Between $4\,t$ and $5\,t$ there is a prominent peak in the \Btwog 
response and a shoulder in the \Boneg response corresponding to transitions between the dispersing portions of the LHB and UHB.  This gives way
to the strong peak in \Boneg symmetry and broad shoulder in \Btwog symmetry at approximately $8\,t$, associated with additional interband scattering pathways
as indicated previously.


\begin{figure}[t]
\centering
\includegraphics[width=0.45\textwidth]{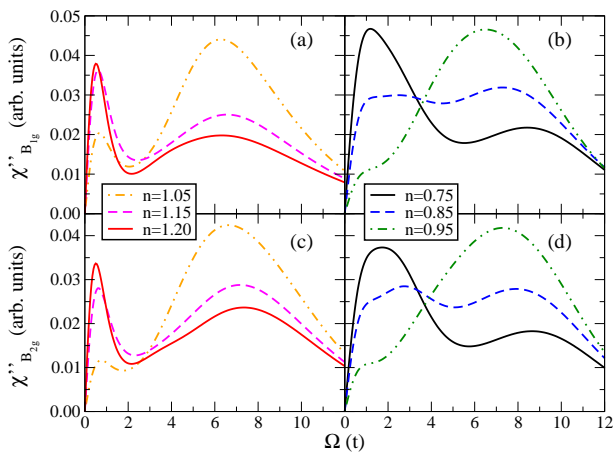}
\caption{(Color online).  Theoretical Raman response in the [(a) and (b)] \Boneg and [(c) and (d)] \Btwog channels for
various doping levels on the [(a) and (c)] electron and [(b) and (d)] hole sides of the phase diagram for the
single-band Hubbard model.}\label{fig:theory-long}
\end{figure}

Figure~\ref{fig:theory-long} shows the \Boneg and \Btwog Raman response for various values of electron [panels (a) and (c)]
and hole [panels (b) and (d)] doping.  Both \Boneg and \Btwog symmetries show progressive transfer of spectral weight to lower energies with doping
away from half-filling as the intraband transition dominates and the single-particle spectral weight is transfered to the QPB.  At low hole-doping ($n=0.95$)
the intraband transition appears as a small shoulder or knee at low energy and the \Boneg and \Btwog response are dominated by the high energy transitions from the LHB to the UHB.  Near optimal doping ($n=0.85$) the Raman response behaves as indicated previously with the low energy response in the \Btwog channel
dominated by the LHB to QPB transition that also gives a very broad peak in \Boneg symmetry.  With additional overdoping ($n=0.75$) the response is
dominated by the intraband transition with the LHB to QPB transition giving an asymmetric shoulder in \Boneg and large broad peak in \Btwog symmetry.  Upon
electron doping, the Raman response in each channel is dominated by the intraband transition at low energy and the LHB to QPB transition at higher energies.  
While the strength of each feature varies with doping, the energy position remains relatively fixed with increasing electron count.


\begin{figure}[t]
\centering
\includegraphics[width=0.45\textwidth]{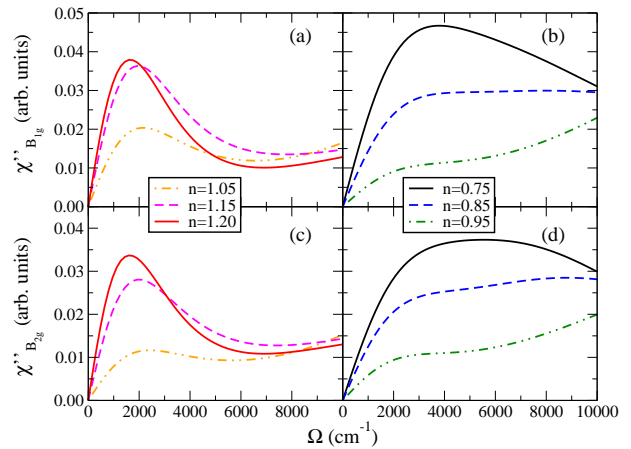}
\caption{(Color online).  Theoretical low-energy Raman response in [(a) and (b)] \Boneg and [(c) and (d)]
\Btwog symmetry from Fig.~\ref{fig:theory-long}.  These results have been converted to cm$^{-1}$
assuming a reasonable value of $t=400$ meV that provides a qualitatively good description of the single-particle spectral function in these systems.}\label{fig:theory-low_energy}
\end{figure}

Figure~\ref{fig:theory-low_energy} shows the Raman response in both channels at low energy associated primarily with the intraband transition.  The
energy scale has been expressed in cm$^{-1}$ assuming a value of $t=400$ meV. For hole doping, the low energy peak grows and appears to shift in energy
from $\sim 2000$ cm$^{-1}$ at low doping ($n=0.95$) to $\sim 4000 - 5000$ cm$^{-1}$ on the overdoped side of the hole-doping phase diagram ($n=0.75$).
However, this apparent shift of energy scale is presumably due to the overlap of several peaks that possibly could be distinguished upon lowering the 
temperature.  With electron doping, the peak intensity grows and the energy shifts to slightly smaller values at the largest electron count ($n=1.20$) near
$\sim 2000$ cm$^{-1}$.  In both cases, upon reducing the temperature from that used in the simulation one expects that the peak will narrow and shift to
even lower energies.

The low energy Raman response, particularly as $\Omega \rightarrow 0$, can be used to obtain an estimate for the effective quasiparticle scattering rate
in each channel and highlight important differences between hole- and electron-doped systems.  This Raman resistivity can be determined from~\cite{RMP}
\begin{equation}
   \Gamma_{\mu}(T) = \left[\frac{\partial \chi^{''}_{\mu}(\Omega,T)}{\partial\Omega}\right]^{-1}_{\Omega = 0}
   \approx \lim_{\Omega \rightarrow 0} \left[\frac{\chi^{''}_{\mu}(\Omega,T)}{\Omega}\right]^{-1}.
   \label{eq:Gamma}
\end{equation}
Figures~\ref{fig:relaxationrates}(c) and (d) show the Raman resistivity in the \Boneg and \Btwog channels for both hole- and electron-doped systems near optimal
doping extracted from the data presented in Fig.~\ref{fig:theory-low_energy}.  As one may expect, the effective scattering rate decreases with decreasing
temperature as well as increasing doping away from half-filling (not shown) on either the hole- or electron-doped sides of the phase diagram indicating that the doped system becomes progressively more metallic.  From Fig.~\ref{fig:theory-low_energy} one also can see that the slope of the Raman response is always larger
for the electron-doped models compared to their counterparts with similar hole doping. This behavior is further reflected in the Raman resistivities shown 
in Figs.~\ref{fig:relaxationrates}(c) and (d).


\section{Experiment}\label{sec:experiments}
\subsection{Samples and experimental details}
Experiments were performed on single crystals of \LSCO (LSCO) and \NCCO (NCCO) on the hole- and electron-doped sides of the phase diagram, respectively. 
The doping level $x$ is indicated in each figure.  Both compounds belong to the 214-family and crystallize in the T'- and T-structure with and without 
oxygen atoms in the apex position, respectively.  The crystals were prepared via the traveling solvent floating zone (TSFZ) technique. The \NCCO samples had 
to be post-annealed in pure Argon to remove the excess oxygen at the apex position and make the samples superconducting.~\cite{Lambacher:2008,Erb:2010}
The annealing protocols for \LSCO are described elsewhere.~\cite{Erb:2010}

\begin{center}
\begin{table*}[t]
\caption[]{Complete list of samples studied partially adapted from Ref.~\onlinecite{Muschler:2010}. The results on La$_{2-x}$Sr$_{x}$CuO$_{4}$ (LSCO) have been published in Ref.~\onlinecite{Muschler:2010}. Those on Nd$_{2-x}$Ce$_{x}$CuO$_{4}$ (NCCO) were taken on freshly prepared single crystals.~\cite{Erb:2010} In the case results similar to ours were published before we give the references in the text and in the figure captions. Samples labeled with $a$ have been prepared by M. Lambacher and A. Erb (WMI Garching),~\cite{Erb:2010} $b$ by Seiki Komiya and Yoichi Ando (CRIEPI, Tokyo  and Osaka University), and $c$ by N. Kikugawa and T. Fujita (Hiroshima and Tokyo). The transition temperatures were measured either resistively or via magnetometry or via the non-linear ac response. The $T_c$ of the $5\%$ sample is the onset point of the transition. $T_N$ was not measured for \LSCO at $x=0.02$ and $0.05$. In the latter case $T_N=0$.}
\centering
\begin{tabular}{c c c c c c c}
 \noalign{\smallskip}\hline\hline\noalign{\smallskip}
 sample &                                        sample ID & doping & ~~~~~$T_c/T_N$ (K) & $\Delta T_c$ (K)&  comment & \\
 \noalign{\smallskip}\hline\noalign{\smallskip}
 ${\rm La_{2} CuO_{4}}$                            & LCO-00     & 0.00 & 0/325 & - & Ar annealed       & $a$ \\
 ${\rm La_{1.98}Sr_{0.02}CuO_{4}}$                 & LSCO-02  & 0.02 & 0/-    & - &       as-grown    & $c$ \\

 ${\rm La_{1.95}Sr_{0.05}CuO_{4}}$                 & LSCO-05    & 0.05 & 5/0   & 3 & O$_2$ annealed & $a$ \\
 ${\rm La_{1.92}Sr_{0.08}CuO_{4}}$                 & LSCO-08    & 0.08 & 18    & 4 & O$_2$ annealed & $c$ \\
 ${\rm La_{1.85}Sr_{0.15}CuO_{4}}$                 & LSCO-15    & 0.15 & 38    & 3 & O$_2$ annealed & $a$ \\
 ${\rm La_{1.83}Sr_{0.17}CuO_{4}}$                 & LSCO-17    & 0.17 & 39    & 1 & O$_2$ annealed & $b$ \\
 ${\rm La_{1.80}Sr_{0.20}CuO_{4}}$                 & LSCO-20    & 0.20 & 24    & 3 & as-grown & $a$ \\
 ${\rm La_{1.75}Sr_{0.26}CuO_{4}}$                 & LSCO-26    & 0.26 & 12    & 3 & O$_2$ annealed & $c$ \\
  \noalign{\smallskip}\hline\noalign{\smallskip}
 ${\rm Nd_{2}CuO_{4}}$                             & NCCO-00    & 0.00 & 0    & - & Ar annealed & $a$ \\
 ${\rm Nd_{1.88}Ce_{0.12}CuO_{4}}$                & NCCO-12    & 0.12 & 0    & - & Ar annealed & $a$ \\
 ${\rm Nd_{1.87}Ce_{0.13}CuO_{4}}$                & NCCO-13    & 0.13 & 9.9    & 7.5 & Ar annealed & $a$ \\

 ${\rm Nd_{1.85}Ce_{0.15}CuO_{4}}$                & NCCO-15    & 0.15 & 23.6    & 1.3 & Ar annealed & $a$ \\
 ${\rm Nd_{1.84}Ce_{0.16}CuO_{4}}$                & NCCO-16    & 0.16 & 16.3    & 2.5 & Ar annealed & $a$ \\

 ${\rm Nd_{1.83}Ce_{0.17}CuO_{4}}$                & NCCO-17    & 0.17 & 5.0    & 3.5 & Ar annealed & $a$ \\
 \noalign{\smallskip}\hline\hline
\end{tabular}
  \label{tab:samples}
\end{table*}
\end{center}

The Raman experiments were performed with a standard light scattering setup.  For excitation an Ar$^+$ laser was used and operated at 
\SIunits{458}{\usk\nano\meter} for \LSCO and \SIunits{514}{\usk\nano\meter} for \NCCO.  The angle of incidence of the exciting light was \SIunits{66}{\degree}.
To achieve proper polarization states inside the sample the polarization of the light outside was controlled with a Soleil-Babinet compensator.  The samples 
were mounted on the cold finger of a He-flow cryostat with temperatures in the range from 4 to \SIunits{330}{\usk\kelvin} and a vacuum of better than 
\SIunits{\microd}{\usk\milli\bbar}.  The scattered light was collected with an objective lens. Photons with selected polarization states were analyzed using 
a Jarrell-Ash 25-100 scanning spectrometer equipped with a CCD camera. The resolution at 458\,nm was 9.5\,cm$^{-1}$ unless otherwise stated. All spectra are 
divided by the thermal Bose factor $\{1+n(\Omega,T)\}=(1-e^{-\Omega/T})^{-1}$ and corrected for the sensitivity of the entire setup including the energy 
dependence of the spectral resolution.

By properly selecting the polarizations of the incident and scattered photons excitations of specific symmetries can be projected out. For particle-hole 
excitations in the cuprates, the \Boneg spectra project mainly the principal axes while the \Btwog spectra contain information about the diagonals of the 
tetragonal Brillouin zone.~\cite{Devereaux:1994} The respective light polarizations and Raman vertices are indicated as insets in 
Fig.~\ref{fig:exp-LSCO_v_NCCO}.


\subsection{Results}\label{sec:results}
\begin{figure}[t]
\centering
\includegraphics[width=0.75\columnwidth]{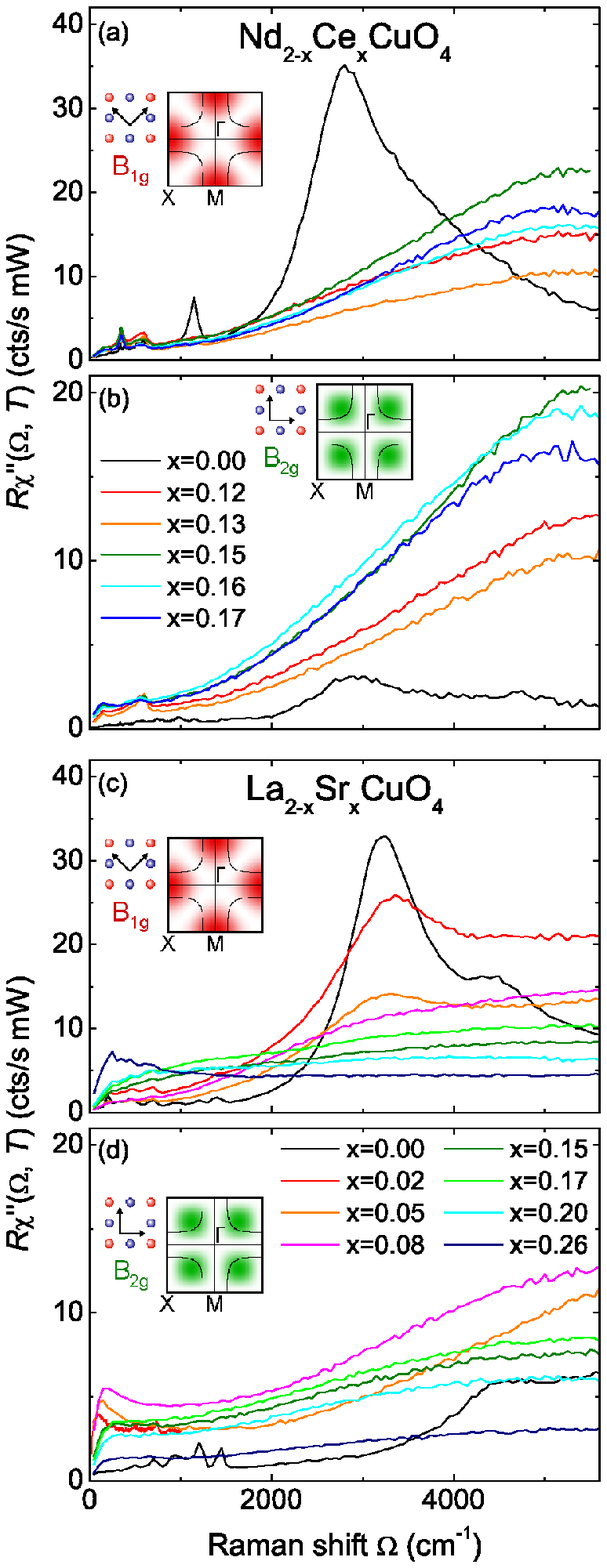}
\caption[]{(Color online). High-energy Raman response of [(a),(b)] NCCO and [(c),(d)] LSCO for electron and hole doping, respectively. The spectra are shown in 
[(a),(c)] \Boneg and [(b),(d)] \Btwog symmetry at various doping levels at temperatures of roughly \SIunits{200}{\usk\kelvin}. The peaks at low doping in (a) 
and (c) are due to two magnon scattering. In general with increasing doping level the spectra on the hole doped side lose intensity while the intensity of the 
spectra on the electron doped side increases.  Part of the results are similar to those of other authors~\cite{Sulewski,Sugai:1988,Sugai:1991,Onose:2004} or were already published in Ref.~\onlinecite{Muschler:2010}.}
\label{fig:exp-LSCO_v_NCCO}
\end{figure}

Fig.~\ref{fig:exp-LSCO_v_NCCO} shows the experimental \Boneg and \Btwog Raman spectra of \NCCO and \LSCO for energies up to 
\SIunits{5600}{\usk\centi\reciprocal\meter} at \SIunits{200}{\usk\kelvin}.  For undoped NCCO ($x=0.00$), the most prominent peak in \Boneg symmetry 
[Fig.~\ref{fig:exp-LSCO_v_NCCO}(a)] is observed at  \SIunits{2900}{\usk\centi\reciprocal\meter} and originates from nearest-neighbor spin flip excitations~\cite{Loudon-Fleury,Sugai:1988,Sulewski,Sugai:1991,Muschler:2010} 
which are not part of the theoretical description in Section~\ref{sec:theory}.  At 1200\,cm$^{-1}$ there is a weak band that originates from two-phonon 
scattering.~\cite{Sugai:2004} Since the resolution is \SIunits{28}{\usk\centi\reciprocal\meter} close to the laserline at \SIunits{514}{\usk\nano\meter} most 
of the phonon excitations are hardly visible.  With doping the two-magnon excitation is suppressed rapidly but traces thereof may still be present at $x=0.12$ (see also Ref.~\onlinecite{Onose:2004}).  Below \SIunits{2000}{\usk\centi\reciprocal\meter} there is little change of the continuum with doping.  At high energies the intensities of the spectra do not depend in a systematic way on doping in contrast to what is observed for YBa$_2$Cu$_3$O$_{6+x}$ (Y-123)~\cite{Tassini:2008} and also in LSCO as shown below. We assume a strong contribution from luminescence, which can be seen by comparing annealed and as grown samples (for a discussion see \emph{e.g.} Ref.~\onlinecite{Muschler:2010}), due to charge traps in the rather imperfect Nd-Ce-O layers which mask the intrinsic effects of the carrier dynamics.

In \Btwog symmetry [Fig.~\ref{fig:exp-LSCO_v_NCCO}(b)] the response below approximately \SIunits{2000}{\usk\centi\reciprocal\meter} is weak for the undoped 
compound presumably due to the small carrier concentration.  The peak at \SIunits{2900}{\usk\centi\reciprocal\meter} does not originate from polarization 
leakage otherwise the two-phonon excitation at 1200\,cm$^{-1}$ present in \Boneg symmetry would be strong enough to be visible as well.  The energy of the \Btwog feature with respect to that in \Boneg symmetry is smaller than in LSCO.  Since the next-nearest neighbor coupling $J^{\prime}$ determines by and large the peak energy in \Btwog symmetry~\cite{Sulewski} we conclude it is weaker in NCCO than in hole-doped systems. The two-magnon scattering in \Btwog symmetry disappears faster with doping than in \Boneg symmetry.  With increasing doping level the spectra gain intensity in the low as well as in the high energy range without significantly changing the spectral shape.


In contrast, a strong doping dependence of the spectral shape is found for LSCO as shown in Fig.~\ref{fig:exp-LSCO_v_NCCO}(c) and (d).
The pronounced \Boneg peak at \SIunits{3300}{\usk\centi\reciprocal\meter} observed at $x=0$ 
corresponds to two-magnon scattering and is progressively 
suppressed upon doping.  The peak at 4200\,cm$^{-1}$ which appears in \Boneg and \Btwog symmetry with the same intensity indicating its \Atwog nature comes from higher order spin excitations including cyclic exchange of spins.~\cite{Vernay:2007}
The low energy response is weak but picks-up intensity with doping. Here, it appears as if spectral weight would be transferred from high to low energies.
In the high energy part the intensity increases between $x=0$
and $0.02$ then decreases monotonically with doping.  This results in a fairly flat spectrum 
at high energy for the highest doping levels.  Above $x=0.2$ 
there appears a peak at low energy which is related to long-lived particle-hole excitations. 
This transfer of intensity is qualitatively predicted already in the Falicov-Kimball and Hubbard models.~\cite{Freericks_1,Freericks_2,Freericks_3,Freericks_4} A discussion in terms of a Fermi-liquid approach is presented elsewhere.~\cite{Prestel:2010}

In \Btwog symmetry, there are pronounced phonon bands below 1000\,cm$^{-1}$ in the undoped compound [Fig.~\ref{fig:exp-LSCO_v_NCCO}(d)] which disappear 
quickly with doping.  In contrast to the \Boneg channel there is no redistribution of spectral weight from high to low energies upon doping. The overall 
variation of the intensity is also non-monotonic. For the highest doping levels the continuum hardly depends on energy.


\begin{figure}[t]
\centering
\includegraphics[width=0.75\columnwidth]{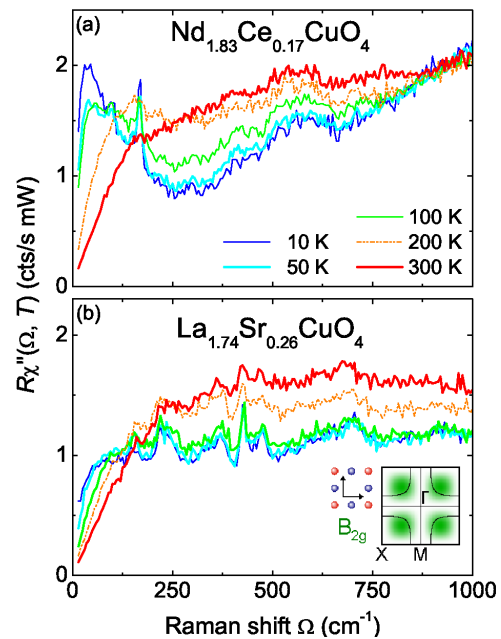}
\caption{(Color online). Temperature dependence of the low-energy Raman response of overdoped (a) NCCO and (b) LSCO in \Btwog symmetry. Only for the electron 
doped compound does there develop a quasiparticle peak at low temperatures.  Results for NCCO with $x=0.15$ displaying similar properties were published in Ref.~\onlinecite{Koitzsch:2003}.  The data for LSCO are adapted from Ref.~\onlinecite{Muschler:2010}.}\label{fig:NCCOLSCOB2g}
\end{figure}

It is particularly instructive to compare the temperature dependence at low energies of electron- and hole-doped materials.  The \Btwog Raman response of 
overdoped NCCO and LSCO are plotted in Fig.~\ref{fig:NCCOLSCOB2g}(a) and (b). In either case the initial slope increases upon cooling corresponding to 
metallic behavior, however, with a significant difference. In NCCO ($x=0.17$) an isolated peak appears at low energy and temperature accompanied by a 
suppression of spectral weight in the range between 200 and 800\,cm$^{-1}$ as already observed earlier by Koitzsch {\it et al}.~\cite{Koitzsch:2003}
In overdoped LSCO ($x=0.26$) there is only an overall reduction of intensity in the entire range without any pile-up at low energies (see also Ref.~\onlinecite{Muschler:2010}). At temperatures above 200\,K the spectral shapes become similar on both sides of half filling.


\section{Discussion}\label{sec:discussion}
\subsection{Overall Features}
It is clear from the theoretical results presented in Section~\ref{sec:theory} and the experimental results presented in Section~\ref{sec:experiments} that theory and experiment cannot be compared quantitatively.  However, a number of qualitative comparisons can be made pertaining to both high and low energy behavior.  The transfer of spectral weight from high to low energies is clearly predicted for the hole doped systems [$n<1$, Fig.~\ref{fig:theory-long}(b) and (d) compared with Fig.~\ref{fig:exp-LSCO_v_NCCO}(c) and (d)]. 
We note that this does not pertain to the two-magnon peak, which is not included in the theory, but rather to the systematic overall increase with hole doping over energies up to at least 1\,eV for $p>0.05$ ($n<0.95$). On the electron-doped side for NCCO, the redistribution is considerably weaker in the theoretical prediction [Fig.~\ref{fig:theory-long}(a) and (c)].  In the experiment [Fig.~\ref{fig:exp-LSCO_v_NCCO}(a) and (b)] there is little change at low energy similar to the theoretical prediction, but in opposition to theory an increase at higher energies which is most likely originating from luminescence as outlined in Sec.~\ref{sec:results}. A significant enhancement of spectral weight is observed at low energies in \Btwog symmetry as predicted in the theoretical results.

The derivation of the exact energy dependence of the electron-hole continuum is experimentally challenging since various processes may contribute. Among the intrinsic contributions are luminescence, resonance enhancement, and spin excitations. There are various studies at relatively low Raman shifts on the dependence of excitations on laser photon energy \cite{Heyen:1990,Ambrosch:2002,Budelmann:2005} whereas there is less material on the high-energy continua. Studies of Kang {\it et al.}~\cite{Kang:PRL77p4434} and Blumberg {\it et al.}~\cite{Blumberg:PRL88e107002} show that the influence of resonances on the low-energy electronic part are mild for blue-green excitation. These are the wavelengths used in the experimental portion of this study. In addition, resonances do not affect the form factors directly since they are dictated by symmetry, but rather affect the relative intensities of the channels and their possible dependence on the excitation energy. However, a calculation of the full vertices is certainly beyond the scope of the present paper and we confine our argumentation to the symmetry part of the vertices.

At low energy, as one can see in Fig.~\ref{fig:theory-low_energy}, the Hubbard model predicts relatively flat spectra on the hole-doped side [panels (b) and (d)] and well-defined peaks for electron doping [panels (a) and (c)]. Since the temperature in the simulations is high, the peaks are wide; however, for the electron doped systems these peaks do originate from 
quasiparticle-like, intraband particle-hole excitations in an almost normal metallic band (see Fig.~\ref{fig:e-doped_disp}). Somewhat differently in the experiment, the peak at low energy appears only at relatively low temperatures. At the moment we do not know what kind of interactions lead to a reduction of the carrier lifetime already around room temperature. Empirically, the possibility exists that the relatively large coefficient $A$ in front of the $T^2$ term of the resisitivity (found at least for LSCO~\cite{Nakamae:2003}) leads to a rapid suppression of the coherence peak with increasing temperature.

Why do we believe that this low-energy peak and the spectral weight suppression originate from particle-hole excitations and not from fluctuations and a pseudogap such as in Y-123~\cite{Tassini:2008} and LSCO~\cite{Tassini:2005,Caprara:2005}?  Simulations~\cite{Prestel:2010} show that the low-energy peak and the dip can in fact result from a reduced quasiparticle damping in a Fermi liquid phenomenology. In addition, the initial slope of the spectra [Eq.~(\ref{eq:Gamma})] follows the resistivity, at least qualitatively, as shown in Fig.~\ref{fig:relaxationrates}(a) while the case is opposite for LSCO [see the \Boneg response in Fig.~\ref{fig:relaxationrates}(b)] and Y-123.

As previously noted, the spectra of materials close to half filling corresponding to $n \sim 1$ or $x \sim 0$ have prominent peaks from magnetic excitations which are not reproduced by the current theory as only the lowest order approximation, i.e. the bare bubble, has been used for the calculation of the Raman response.~\cite{Loudon-Fleury,Shraiman-Shastry}  
Systematic studies in Y-123, Bi$_2$Sr$_2$CaCu$_2$O$_{8+\delta}$ (Bi-2212), Tl$_2$Ba$_2$CuO$_{6+\delta}$ (Tl-2201), and LSCO show~\cite{Muschler:2010} that this approximation is too simple, in particular at lower doping. As demonstrated in Bi-2212 the collapse of the approximation appears to set in rather abruptly at $n \sim 0.79$ ($p \sim 0.21$) and has no direct correspondence in the single-particle spectral function.~\cite{Freericks_3}  

How might one expect the overall features to differ under a scenario in which the Hubbard $U$ were allowed to decrease significantly with doping away from half-filling on either the hole- or electron-doped sides of the phase diagram?  One expects a more pronounced transfer of spectral weight from high to low energies than that observed in Fig.~\ref{fig:theory-long} together with a pronounced shift of residual high energy spectral weight to lower energies corresponding to the reduction in Hubbard $U$.  Doping-dependent $U$ also would lead to the appearance of sharper, more metallic quasiparticle peaks at low energies especially on the hole-doped side of the phase diagram at greater variance with experimental observations compared to the current theoretical analysis.  While one can argue that vertex corrections may tend to reduce or flatten the Raman response at low energies and introduce features corresponding to magnetic excitations lacking in the lowest order approximation used in this study, the effect of these corrections with a doping-dependent $U$ would tend to decrease significantly with doping and potentially suppress magnetic excitations in the response more rapidly than indicated by experimental observations.       


\subsection{Relaxation Rates}
\begin{figure}[t]
\centering
\includegraphics[width=0.45\textwidth]{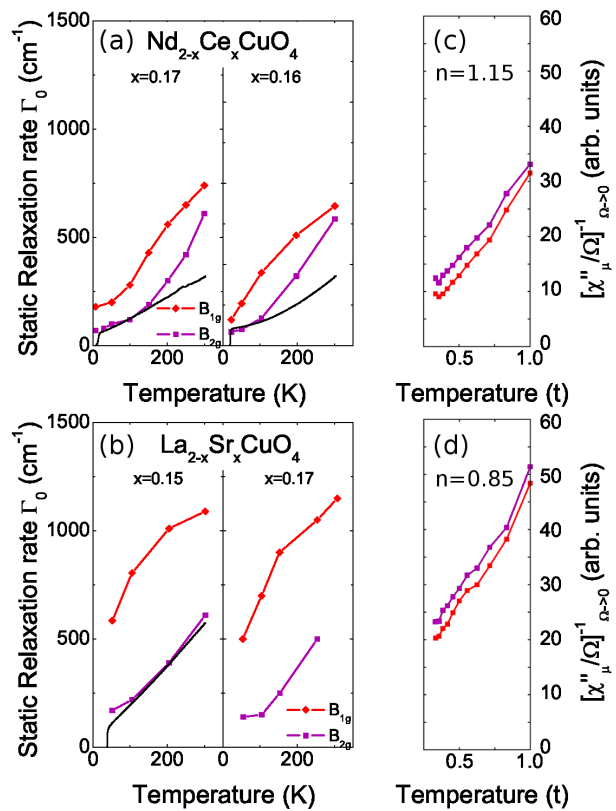}
\caption{(Color online). Temperature dependence of the (a) and (b) experimental and (c) and (d) theoretical Raman resistivities in \Boneg (red) and \Btwog (purple) symmetry obtained from the electron (\NCCO) and hole-doped (\LSCO) samples around optimal doping, respectively.}\label{fig:relaxationrates}
\end{figure}

The method for extracting Raman relaxation rates from the spectra has been described elsewhere.~\cite{Opel:2000}  The calculated Raman relaxation rates from 
the experimental spectra are shown in Figs.~\ref{fig:relaxationrates}~(a) and (b) in addition to the inverse initial slopes of the theoretical spectra as determined via Eq.~(\ref{eq:Gamma}) as shown in Figs.~\ref{fig:relaxationrates}~(c) and (d).  For both samples the experimental Raman relaxation rates are higher in the \Boneg than in the \Btwog channel.  The relaxation rates in \Btwog have a similar temperature dependence for the electron as well as for the hole doped side showing a concave curve following approximately $T^{\alpha}$ with $1 < \alpha < 2$.


On the hole-doped side the relaxation rates in \Boneg have a convex form ($\alpha < 1$).  This is different from the electron doped side.  Here the form 
changes from convex to concave for a small variation in doping from $x=0.16$ to $0.17$. This is actually the range of doping where a crossover between a 
small and a large Fermi surface is observed with quantum oscillations.~\cite{Helm:2009}  The inverse initial slopes of the theoretical spectra are different 
in that the relaxation rates in \Btwog are bigger than in \Boneg for all temperatures on both sides of the phase diagram.  Additionally the temperature 
dependence is concave for the electron as well as for the hole doped side.  However, what can be qualitatively resolved, in agreement with experiments, is the fact that the relaxation rates for both symmetries are smaller on the electron than on the hole doped side of the phase diagram.

The huge difference in the experimental relaxation rates between the \Boneg and \Btwog response in LSCO is not yet fully understood. As mentioned above it appears in a doping range in which the dichotomy between nodal and anti-nodal quasiparticles is still small~\cite{ARPES_RMP} while the Raman spectra exhibit an abrupt onset of strong relaxation close to $(\pi,0)$.~\cite{Freericks_3}  
The discrepancy appears to indicate that Raman vertices should be renormalized at least at low doping.  The observed theoretical behavior for hole-doping is in better qualitative agreement with the experimental results on overdoped \LSCO (not shown) where experimentally one observes more Fermi liquid-like behavior with metallic quasiparticles in the normal state as opposed to the strange metal and pseudogap phases at doping levels closer to half-filling.  At the accessible temperatures in the theoretical analysis, there is no indication of a pseudogap in the single-particle spectra; however, from other small cluster studies one anticipates the formation of a pesudogap at considerably lower temperatures.~\cite{Maier_Quantum_Cluster_RMP}  The pseudogap manifests as a reduction in single-particle spectral weight near the $(\pi,0)$ points and one anticipates a concomitant reduction in the \Boneg Raman response and significant increase in the corresponding relaxation rate potentially bringing the theoretical and experimental results into better agreement.  Lower temperatures also tend to sharpen quasiparticle features potentially reducing relaxation rates, especially in the \Btwog channel extracted from quasiparticles near $\sim (\pi/2,\pi/2)$ that are particularly wide in the current simulations.  
  

\subsection{Evolution of the spectral weight with doping}
Finally, we analyze the ratio of the Raman intensities in the \Boneg and \Btwog channel $I_{\mathrm{B1g}}/I_{\mathrm{B2g}}$.  For deriving $I_{\mathrm{B1g}}/I_{\mathrm{B2g}}$ we integrate the experimental intensity in \Boneg and \Btwog channels over the range 800-\SIunits{1000}{\usk\centi\reciprocal\meter}, labeled $I_{\mathrm{B1g}}$ and $I_{\mathrm{B2g}}$, respectively.  Over this range the experimental spectra are fairly temperature 
independent.~\cite{Opel:2000}  For a simple tight-binding band structure, one expects this ratio to be given by $(t/2t^{\prime})^2$; however, discrepancies between the experimental response and this simple expectation, at least on the hole-doped side of the phase diagram, were pointed out in earlier studies.~\cite{Katsufuji:1993,Naeini:2000}  While a similar analysis of integrated intensity from the theoretical results is complicated by the lack of vertex corrections and the relatively high simulation temperature leading to significant low energy peaks in the response rather than flat, featureless spectra in the indicated energy range (see Fig.~\ref{fig:theory-low_energy}), we perform a corresponding analysis of the integrated spectral weight as a function of doping and temperature over the same energy range (see Fig.~\ref{fig:theory-low_energy}) which at least removes a degree of arbitrariness from the comparison between experiment and theory.    


\begin{figure}[t]
\centering
\includegraphics[width=0.45\textwidth]{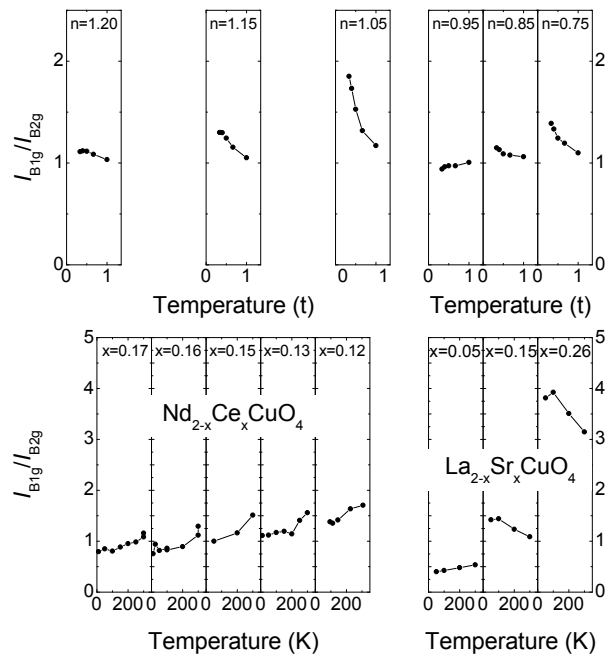}
\caption{Ratio of the intensities in $I_{\mathrm{B1g}}/I_{\mathrm{B2g}}$ as extracted from the theoretical (upper) and experimental (lower) 
Raman spectra for electron and hole doping levels indicated in each panel.}\label{fig:ratio}
\end{figure}

Fig.~\ref{fig:ratio} shows the ratio $I_{\mathrm{B1g}}/I_{\mathrm{B2g}}$ both calculated from the theoretical response and extracted from the experimental spectra.  On the electron-doped side the experimental ratio is slightly temperature and doping dependent for those samples analyzed near optimal doping.  
While the ratio increases slightly with decreasing doping, the temperature dependence of the ratio remains relatively unchanged, increasing moderately with increasing temperature.  From the theoretical results, we analyze the ratio both for optimal doping as well as for underdoped and overdoped systems.  Close to half-filling, we find that the ratio increases significantly with decreasing temperature although it remains within a factor $\sim 2$ of the value extracted near optimal doping.  While this temperature dependence differs significantly from that found experimentally near optimal doping, it is not completely unexpected given the experimental and theoretical evolution of the Fermi surface and bandstructure with doping.~\cite{NPArmitage}  At low doping levels small, electron-like Fermi pockets first appear near $(\pm\pi,0)$ and $(0,\pm\pi)$ at low temperatures, due to antiferromagnetic folding, which in the simple response calculated here would lead to a significantly larger low energy response in the \Boneg channel compared to the \Btwog channel.  With increasing temperature, one expects increasing response in the \Btwog channel and a reduced ratio due to an increase in spectral weight in the nodal region.  With increased electron doping the spectral weight in the nodal region increases leading to a reduction in the intensity ratio near optimal doping.  The theoretical ratio increases slightly with decreasing temperature, at odds with the experimental result; although it appears to saturate at the lowest simulated temperatures and has a value that compares quite well with that extracted from experiments.  Upon additional electron doping a well-defined hole-like Fermi surface forms, in agreement with experiment; however, the antinodal BZ crossing is pulled progressively away from the $(\pi,0)$ points leading to a reduction in the integrated Raman response in the \Boneg channel with a subsequent reduction in the intensity ratio.  While the temperature dependence of the ratio extracted from the theoretical Raman response is not in quantitative agreement with the experimentally determined intensity ratio, the general reduction in the ratio with increased doping is captured by the theoretical analysis and one also should keep in mind the limitations of the theoretical analysis (lack of vertex corrections) and significant difference in temperature scales.            

On the hole-doped side we can compare data over a much wider doping range.  For low doping, at the onset of superconductivity ($x=0.05$), the experimental ratio increases slightly with increasing temperature, but has a value that remains less than one over the studied temperature range.  Upon hole doping away from half-filling, one observes Fermi arcs experimentally in the single-particle spectral function centered on the nodal points with the appearance of a pseudogap in the antinodal region.~\cite{ARPES_RMP}  The effect of these features would tend to suppress the integrated \Boneg Raman response and hence the intensity ratio within the simple picture used to describe the response theoretically, in apparent agreement with the experimental findings.  With additional hole doping, the Fermi arcs connect as the pseudogap closes in the single-particle spectral function near optimal doping.  Experimentally, the integrated intensity ratio increases by a factor of $\sim 2 - 3$ over its value at underdoping and the ratio increases with decreasing temperature.  Upon overdoping, the antinodal BZ crossings move toward the $(\pi,0)$ points as the van Hove singularity approaches the Fermi level in the same region of the BZ.   This would significantly increase the integrated \Boneg response in the simple band picture used in the theoretical analysis.  Experimentally the intensity ratio for the overdoped system increases substantially with decreasing temperature and has a value $\sim 3$ times larger than that found near optimal doping.  The ratio extracted from the theoretical response agrees qualitatively with the experimental results both in the general trends with hole doping away from half-filling and changes in the ratio with increasing or decreasing temperature.  The lack of quantitative agreement may be attributed to the relatively high simulation temperatures that preclude the appearance of a pseudogap in the single-particle spectral function at low doping and that significantly broaden quasiparticle features irrespective of the doping level and to the lack of vertex corrections that would renormalize the Raman response in each channel, particularly at low doping.           


\section{Conclusions}\label{sec:conclusions}
While it is quite clear that the theoretical and experimental results presented here cannot be compared quantitatively, qualitative comparisons exist that pertain to both the high and low energy Raman response in hole and electron doped systems.  Speaking generally, the theoretical response shows a significant transfer of spectral weight from high to low energies with increased hole doping away from half-filling in agreement with experiment.  With electron doping, this transfer is less pronounced, except in the underdoped regime, also in agreement with experiment.  At lower energies, the model calculations predict relatively flat spectra on the hole-doped side of the phase diagram and well-defined peaks associated with quasiparticle-like excitations for electron-doped systems.  The lack of vertex corrections in the theoretical response means that this general agreement does not apply to the two-magnon peak associated with magnetic excitations that appears prominently in the experiment, but rather to the overall, systematic trends in the data that appear with either electron or hole doping and already reveal an electron/hole doping asymmetry.  While similar overall trends would be expected from a theoretical analysis including significant reductions in electron correlations with doping away from half-filling (doping-dependent Hubbard $U$), the effect would be more pronounced and at greater variance with the experimental observations.  In particular, one expects more metallic behavior at even lower electron or hole doping and would anticipate significantly reduced influence from vertex corrections with doping that would be necessary to bring the theoretical and experimental observations into better qualitative and quantitative agreement with one another.

Additional electron/hole doping asymmetry has been observed experimentally in both the Raman relaxation rates in the \Boneg and \Btwog channels as well as in the ratio between the integrated Raman intensities in these two channels.  The Raman relaxation rates or resistivities extracted from the theoretical model near optimal doping reveal a similar asymmetry at least in the \Boneg channel.  As argued, the inclusion of pseudogap behavior in the antinodal region and sharper quasiparticle-like features in the near nodal region with reduced temperatures may bring the theoretical and experimental results into better qualitative agreement, especially concerning the behavior of the Raman resistivity in the \Btwog channel where there is a less pronounced asymmetry between electron and hole doping.  

The experimentally observed ratio between the integrated intensities in the \Boneg and \Btwog channel has been shown systematically over a wide range of hole doping and temperature and near optimal doping over a wide range of temperatures on the electron-doped side of the phase diagram.  On the hole-doped side of the phase diagram, the ratio extracted from the theoretical model agrees qualitatively with the experimental results both as a function of doping and temperature, although the effective temperatures differ significantly.  The relatively high simulation temperatures preclude the appearance of a pseudogap in the single-particle spectral function at low doping and broaden quasiparticle features, especially at high doping.  Vertex corrections, missing from the theoretical analysis, would significantly renormalize the Raman response in each channel, particularly at low doping.  Incorporating both of these corrections would improve the quantitative agreement between the theoretical and experimental ratios in hole-doped systems. 

On the electron-doped side of the phase diagram near optimal doping the experimental intensity ratio is only slightly temperature and doping dependent.  
From the theoretical model we have analyzed the integrated intensity ratio over a wider range of electron doping and predict a significant difference between the electron and hole doping evolution.  Close to half-filling, the theoretical ratio increases significantly with decreasing temperature, showing similar behavior to the theoretical and experimental ratio observed for hole-overdoping.  With additional electron doping the ratio progressively decreases.  The doping evolution of the theoretical ratio can be partially understood by considering the evolution of the single-particle spectral function and the Fermi surface.~\cite{ARPES_RMP,NPArmitage} While the Fermi surface consists of small Fermi arcs with a significant pseudogap on the hole-underpoded side of the phase diagram leading to a small integrated intensity ratio, the Fermi surface consists of small electron pockets on the electron-underdoped side of the phase diagram leading to a rather large integrated intensity ratio.  The evolution of the Fermi surface and bandstructure with additional doping on either side of the phase diagram appears to correlate quite well with the doping evolution and the electron/hole asymmetry observed in the intensity ratio.  

While the general trends observed in the theoretical and experimental results agree qualitatively and reveal significant electron/hole doping asymmetries in several quantities, one should keep in mind the limitations of the theoretical approach:  (1) the simulation temperatures are quite high when compared to the experiments meaning that effects from features like the pseudogap that manifest at much lower energies are neglected, (2) the theoretical analysis based on the lowest order approximation to the Raman response neglects important vertex corrections that renormalize the response, especially at low energy, and therefore precludes the appearance of magnetic excitations that are prominent in the experiment.  Although in general the results are encouraging, further improvements including materials specificity missing from the current theoretical analysis and the inclusion of vertex corrections by a direct determination of the Raman response, typical for the charge and spin response functions (charge and spin dynamical structure factors), would certainly improve agreement.  Hence, apart from some qualifications explicitly stated, the Hubbard model is found also here to capture the essential physics of the cuprates around half-filling.

\section{Acknowledgements}
B.Mo., S.J. and T.P.D. acknowledge support form the U.S. Department of Energy, Office of Basic Energy Sciences, Materials Sciences and Engineering Division, under Contract No. DE-AC02-76SF00515.  B. Mu., W.P., R.H., M.L. and A.E. acknowledge support from the Deutsche Forschungsgemeinschaft (DFG) via Research Unit FOR~538 (Grants No. Ha2071/3 and Er342/1). B.Mu. and R.H. gratefully acknowledge support by the 
Bavarian Californian Technology Center (BaCaTeC).  S.J. acknowledges support from Foundation for Fundamental Research on Matter.

\bibliographystyle{apsrev}
\bibliography{Moritz_Raman_9_2011}

\end{document}